# Electronic Nose for Agricultural Grain Pest Detection, Identification, and Monitoring: A Review

*Chetan M Badgujar[1*], Sai Swaminathan[2], and Alison Gerken[3]*

[1]*Department of Biosystems Engineering & Soil Science, University of Tennessee, Knoxville, TN, USA.*

[2]*Department of Electrical Engineering & Computer Science, University of Tennessee, Knoxville, TN, USA.*

[3]*USDA-ARS, Center of Grain & Animal Health Research, Manhattan, KS, USA.*

**Highlights:**

- A systematic literature review was conducted on 21 research studies.
- E-nose technology is low-cost, rapid, non-invasive, and accurate for odor-based pest detection.
- E-nose can detect and identify microscopic (Fungi) and hidden insects with good accuracy.
- E-nose performance is influenced by storage duration, storage parameters, pest species type, and pest density.

**Abstract:** Biotic pest attacks and infestations are major causes of stored grain losses, leading to significant food and economic losses. Conventional, manual, sampling-based pest recognition methods are labor-intensive, time-consuming, costly, require expertise, and may not even detect hidden infestations. In recent years, the electronic nose (e-nose) approach has emerged as a potential alternative for agricultural grain pest recognition and monitoring. An e-nose mimics human olfactory systems by integrating a sensor array, data acquisition, and analysis for recognizing grain pests by analyzing volatile organic compounds (VOCs) emitted by grain and pests. However, well-documented, curated, and synthesized literature on the use of e-nose technology for grain pest detection is lacking. Therefore, this systematic literature review provides a comprehensive overview of the current state-of-the-art e-nose technology for agricultural grain pest monitoring. The review examines employed sensor technology, targeted pest species type, grain medium, data processing, and pattern recognition techniques. An e-nose is a promising tool that offers a rapid, low-cost, non-destructive solution for detecting, identifying, and monitoring grain pests, including microscopic and hidden insects, with good accuracy. We identified the factors that influence the e-nose performance, which include pest species, storage duration, temperature, moisture content, and pest density. The major challenges include sensor array optimization or selection, large data processing, poor repeatability, and comparability among measurements. An inexpensive and portable e-nose has the potential to help stakeholders and storage managers take timely and data-driven informed actions or decisions to reduce overall food and economic losses.

**Keywords:** Biomimicry, Biosensors, Grain storage, Integrated pest management, Sensor technology.

**Introduction:**

Biotic pests can cause significant losses in grain quantity and quality during grain storage. The estimated pest-driven storage losses are up to 9% in developed nations and around 20% or more in developing nations (Kumar, 2017; Phillips & Throne, 2010). Globally, pest-driven storage losses are estimated to be around $100 billion/ year (Zhu et al., 2022). Protecting grains from pest-driven losses is a critical

[*]Corresponding author (chetanmb@utk.edu)

component of ensuring global food security, nutrition, and the safety of a continuously growing population (Phillips & Throne, 2010). Excessive pest populations in specific geographic locations can result in quarantine restrictions, leading to import or export restrictions or bans, further challenging the global food supply chain (Adler et al., 2022; Kumar, 2017). Therefore, it is important to reduce overall pest-driven grain storage losses.

Integrated pest management (IPM) strategies have been recommended as a holistic approach to pest management to reduce overall pest-driven losses (Badgujar et al., 2023b; Zhu et al., 2022). The success of the IPM program largely depends on early pest detection, identification, and establishing pest threshold via an active and timely monitoring approach (Hagstrum, 2012). The conventional insect detection and identification methods include grain sampling, visual inspection, sieving, insect traps, and grain probes (Flinn & Hagstrum, 2014; Hagstrum, 2012). These methods are accurate and reliable but have certain limitations, including being destructive, labor-intensive, and time-consuming, and often require large sampling, specialized tools/equipment, or expertise (Badgujar et al., 2023b; Flinn & Hagstrum, 2014; Hagstrum, 2012; Hagstrum & Athanassiou, 2019). In addition, these methods may not even detect internal feeder or grain infestations. Sensor-based pest detection methods, including acoustic (Mankin et al., 2021; Mankin et al., 2011), imaging (Badgujar et al., 2023b) , and near-infrared (NIR) spectroscopy (Liu et al., 2009), coupled with data-driven modeling (Santiago et al., 2017), is being actively developed and explored in recent years. These methods offer several advantages, such as being non-destructive, rapid, and automated (Liu et al., 2009; Mankin et al., 2011; Santiago et al., 2017). However, image and NIR spectroscopy-based systems are generally limited to detecting or identifying insects on grain surfaces alone, or through probe and trap captures, and often fail to detect hidden or internal infestations typically found in bulk grain storage (Badgujar et al., 2023a, 2023b; Liu et al., 2009; Ni et al., 2024).  On the other hand, acoustic-based systems face limitations such as interference from external and environmental noise, which can affect detection accuracy (Mankin et al., 2021; Mankin et al., 2011; Santiago et al., 2017). Therefore, there is significant interest in developing methods that can overcome the limitations of current sensor-based approaches, with the capability to accurately detect and identify pest-driven infestations or spoilage at an early stage.

A biomimicry-based approach called "electronic nose" (e-nose) has been explored for detecting and identifying specific odors or volatile organic compounds (VOCs) associated with insect pests. E-nose imitates the human sense of smell by using sensor arrays, signal processing techniques, and data-driven models to detect and identify associated odors. E-noses have been commonly used for odor recognition in application areas not limited to human or animal health (Wilson, 2018, 2023), environmental monitoring (Capelli et al., 2014), crop disease/pest (Cui et al., 2018; Fundurulic et al., 2023), food control and quality (Loutfi et al., 2015; Tan & Xu, 2020), cosmetics (Wilson & Baietto, 2009), pharmaceutical (Wasilewski et al., 2019), and security (Macías-Quijas et al., 2019). In recent years, an e-nose application has also been explored for detecting pest infestations in stored grain. The e-nose technology offers rapid, non-destructive, and automated tools, with the potential to recognize hidden or covered insects, particularly in bulk grain storage (Mishra et al., 2018b; Srivastava et al., 2019a).  It may overcome the collective limitation of sensor-based methods (i.e., imaging, NIR-spectroscopy, acoustics), making it more suitable for pest recognition in a variety of bulk grain storage environments.

The increased global grain production and storage demands require non-invasive, real-time, and automated pest detection tools to safeguard food production (Flinn & Hagstrum, 2014; Hagstrum &

Athanassiou, 2019; Phillips & Throne, 2010). E-nose technology shows promise in addressing these needs, though its potential remains underutilized due to a lack of well-organized and curated literature resources. The e-nose pest recognition performance is primarily influenced by sensor type, microcircuitry design, software integration, data processing methods, and classification models (Tan & Xu, 2020; Wilson & Baietto, 2009). For example, a machine learning (ML) or artificial intelligence (AI) approach has significantly enhanced e-nose performance (Li & Xu, 2014; Srivastava et al., 2019b; Ying et al., 2015). However, variations in sensor type, design, and integration of AI/ML with e-nose systems for pest detection are yet to be documented or critically reviewed. Moreover, pest species, storage practices, and storage conditions vary from region to region and crop to crop. This variation would influence e-nose performance. Thus, this study categorizes findings based on geography, grain type, or storage environment, to provide valuable insights for localized implementation.

Overall, several studies have explored e-nose applications for pest detection in stored grains, but the existing knowledge is fragmented and dispersed across different grain types, pests, and methodologies. A comprehensive review is lacking that synthesizes these findings to provide a cohesive understanding of trends, advancements, and limitations. This review aims to bridge this gap by documenting the existing literature with well-organized and curated resources while highlighting key advancements, evaluating modeling strategies, and identifying future directions for effective and scalable e-nose applications in stored grain pest management.

**Materials and methods:**

Overview of electronic nose technology: The e-nose is a bio-inspired sensing system designed to mimic the human olfactory system by detecting and discriminating between complex odors through patterns of VOCs. Their core functionality lies in their ability to capture, transduce, and interpret the unique chemical signatures released by various substances in the gas phase. The major components involved in e-nose operation and working principle are presented in Figure 1.

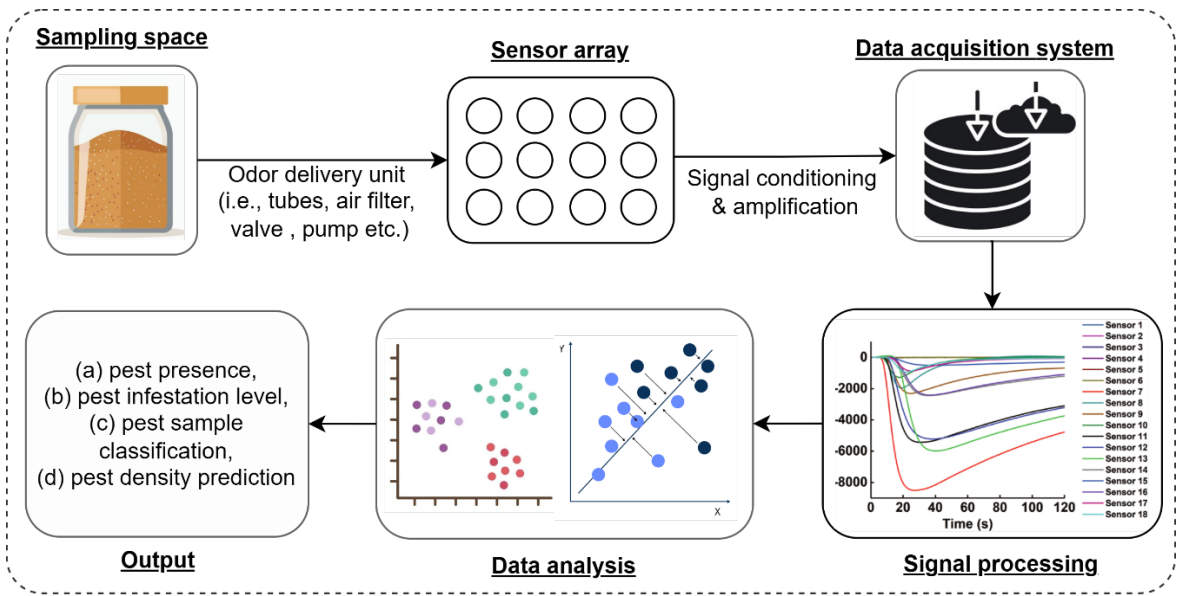

*Figure 1: Working principle of e-nose along with its major components.*

The process begins with collecting VOC from the sampling space (i.e., headspace) using an odor delivery unit and transferring it to the sensor array. The sensor array is composed of chemically sensitive materials that respond to the presence of different VOC. A commonly employed sensor type in modern e-nose systems is the metal-oxide semiconductor (MOS) and conducting polymer (CP) sensors (Wilson & Baietto, 2009). These sensors exhibit changes in electrical conductivity in response to chemical interactions with nearby gas or VOC molecules. These interactions are nonspecific and complex, allowing the sensors to detect a broad range of volatile compounds, though with differing sensitivities and selectivity (Tan & Xu, 2020). The generated electrical signals by multiple sensors are conditioned, amplified, and stored in a data acquisition system for analysis. This diversity in sensor response generates a multidimensional signal pattern, often referred to as a "smell print," which can be used to distinguish between different odor sources.

The sensor responses are often time-dependent, capturing the transient dynamics of gas-sensor interactions. For signal processing and analysis, feature vectors are typically extracted from short time windows (e.g., 0.5 seconds or 1-2 min in some cases), capturing statistical features such as the mean, standard deviation, and time derivatives. These vectors are then used for multivariate data analysis. In data analysis, various learning techniques such as principal component analysis (PCA), discriminant function analysis, cluster analysis, and machine learning algorithms (i.e., k-means clustering and artificial neural networks) are commonly applied to classify and interpret the sensor data. In summary, electronic noses represent a sophisticated fusion of material science, embedded electronics, and machine learning. By leveraging multi-sensor arrays and extracting patterns from high-dimensional data, they offer a promising tool for real-time odor monitoring and classification across a wide range of industrial and environmental settings.

<u>Literature review methodology:</u> We adopted a Systematic Literature Review (SLR) approach to collect, analyze, and synthesize existing research literature on e-nose-based pest detection and identification in stored grain environments. The SLR methodology allows extensive and relevant literature coverage with reproducibility (Mengist et al., 2020). The review aims to address the following research questions: (a) What are the most used e-nose sensor types? (b) What insect species or grain types have been studied, and what are their geographical locations? (c) What storage and pest factors influence the e-nose performance? (d) What data analysis methods and models have been employed, and how do they influence detection performance? (e) What are the strengths, challenges, and opportunities associated with e-nose systems in storage environments?

The articles were retrieved from the most common databases (i.e., Scopus, ScienceDirect, and Google Scholar) with the keyword search strategy, which included the following major keywords: "electronic nose," "e-nose," "grain storage," "pest detection," and "insect infestation". The search was conducted in March 2025, which resulted in 188 documents. The document exclusion criteria were established to ensure that only relevant literature was considered, as mentioned in Figure 2. The selection process is illustrated in the PRISMA flow diagram (Page et al., 2021) as shown in Figure 2. A total of 21 articles were selected for SLR based on defined criteria, which seems to be sufficient for SLR (Mota et al., 2021). Each selected article was studied carefully, with key information recorded not limited to publication year, study location, target pest species, grain medium, sensor type, modeling techniques, performance metrics, and key findings. A thematic analysis was conducted to identify trends, challenges, and emerging directions on e-nose research for pest detection in stored grain systems.

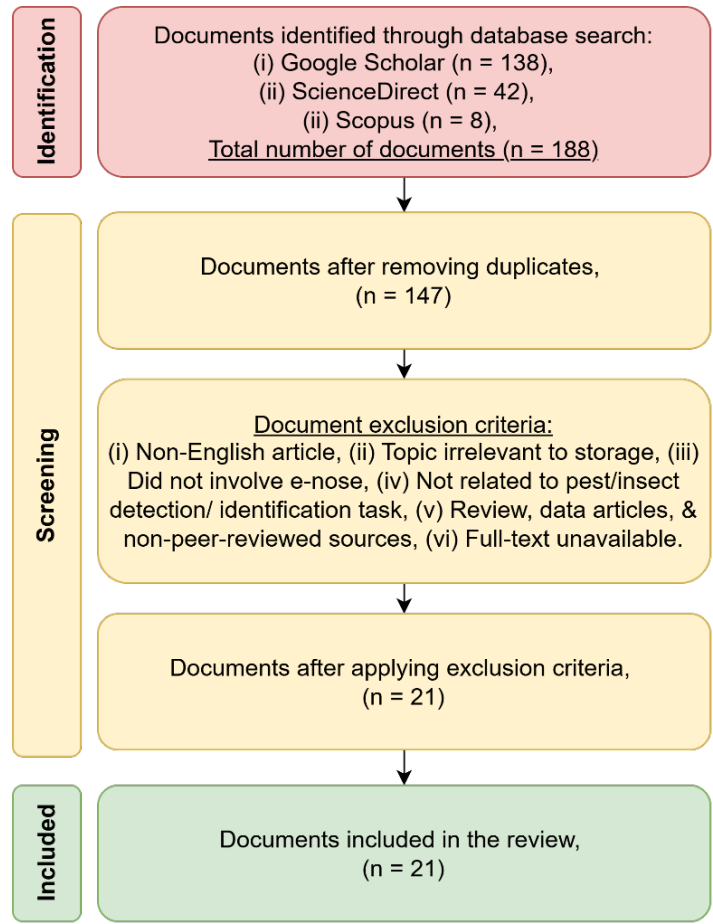

*Figure 2: Overview of employed literature review procedure based on PRISMA flow-chart (Page et al., 2021).*

**Results and discussion:**

The annual trend and global distribution of e-nose published literature are presented in Figure 3. The number of publications remains relatively low (i.e., 2) from the 1999 to 2009 period. However, a noticeable increase in e-nose literature was observed between 2010 to 2025, with a slight decline from 2021-2025, but remains higher than pre-2010. Overall, a significant increase in research interest in e-nose technology has been observed in recent decades. The geographic distribution represented reflects the global research landscape, interest, and research hotspots. The majority of research is conducted in developing Asian countries, dominated by China (6), India (5), Thailand (2), and Iran (1). This might be primarily because the estimated pest-driven storage losses are significantly higher in developing Asian countries (20% or more in losses) than in developed countries (up to 9%), which suggests active and ongoing research efforts to reduce overall storage losses (Phillips & Throne, 2010).

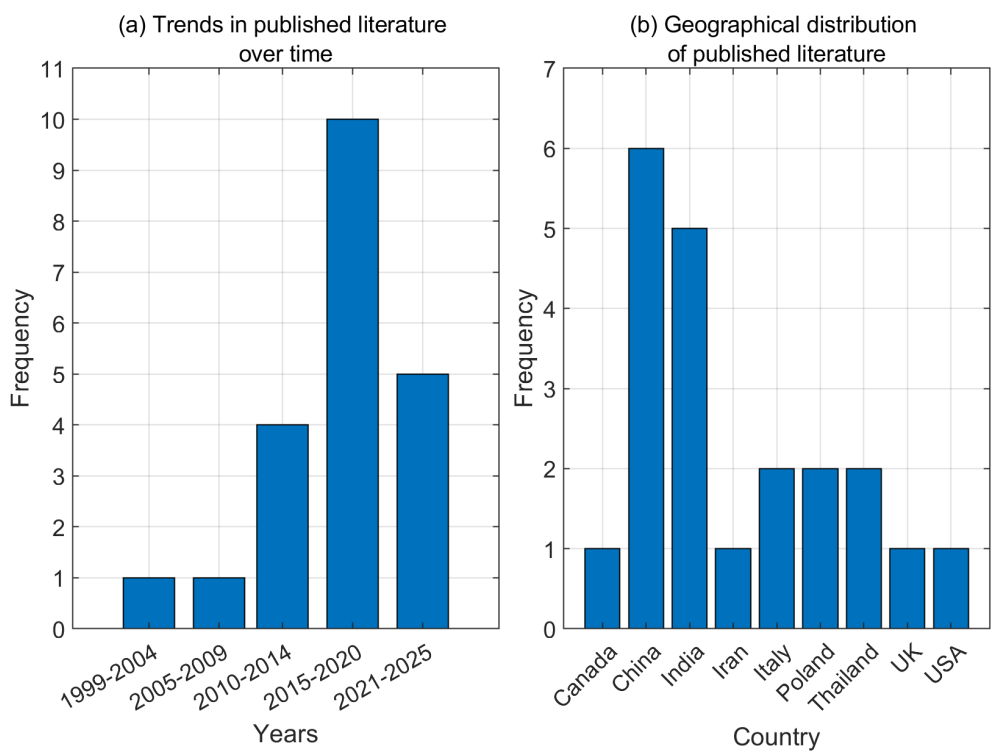

*Figure 3: Electronic nose published literature: (a) trend in literature over time and (b) geographical distribution of published literature.*

A brief summary of the 21 studies that explored the application of e-nose for biotic pest detection, identification, monitoring, and/or quantification of infestation levels are presented in Table 1. The e-nose was employed in a variety of tasks, including detecting mites (Ridgway et al., 1999), insect damage or infestation (Wu et al., 2013; Zhang & Wang, 2007; Zhou et al., 2021), mycotoxins (Leggieri et al., 2022; Lippolis et al., 2014), fungal contamination (Borowik et al., 2024; Gancarz et al., 2017; Jiarpinijnun et al., 2020), classifying pest infestation (Li & Xu, 2014; Srivastava et al., 2019a, 2019b), predicting insect presence (Mishra et al., 2018a; Xu et al., 2017), moldy grains (Ying et al., 2015; Zhang et al., 2022), and classifying or determining infestation level (Mishra et al., 2018b) or pest density (Hou et al., 2025; Nouri et al., 2019). The e-nose success in each task was evaluated by several metrics depending on the specific task goal (Table 1), which included classification accuracy or rate, discrimination index, mean recognition percentage, and regression fitting values (R or $R^2$). Collectively, all reviewed studies demonstrated the satisfactory to excellent performance of e-nose in the respective applications, highlighting its accuracy, reliability, and potential of being a promising tool for pest detection, identification, and monitoring in stored grain environments.

Fungi and molds are microscopic in nature and are not visible to the naked eye while they are growing and damaging grain or food products (USDA, 2016a). Thus, the existing sensor-based approaches, particularly imaging and acoustics-based systems, often fail to detect their presence unless significant growth or damage has already occurred. However, e-nose systems were successful in detecting fungal and mold infestations in grains. For example, Lippolis et al., (2014) and Gancarz et al., (2017) focused on fungal quality assessment in durum wheat and rapeseed, respectively, while Ying et al., (2015) and Jiarpinijnun et al., (2020) achieved high mold recognition accuracies (above 93%) in rice and oats. These studies indicate the potential of e-noses in assessing microbial biotic stressors.

*Table 1: A summary of e-nose applications in storage pest detection, identification, and monitoring.*

| Author | Task performed with an e-nose | Major findings |
|---|---|---|
| (Ridgway et al., 1999) | Detection of mite infestation in wheat. | Achieved classification accuracy of over 83%. |
| (Zhang & Wang, 2007) | Detecting wheat sample age and insect damage with e-nose | E-nose discriminates successfully among wheat of different ages and with different degrees of insect damage. |
| (Deshpandey et al., 2010) | Sense grain storage conditions including pest damage | E-nose successfully differentiates grain samples at different physical conditions including pest damage |
| (Wu et al., 2013) | Detect insect infestation in wheat | E-nose detection capabilities were influenced by grain moisture, infestation level, and insect species. |
| (Li & Xu, 2014) | Proposed new modeling approach using olfactory neural network and SVM with e-nose to classify pest infestation in wheat | Achieved 100% classification rate |
| (Lippolis et al., 2014) | Screening of mycotoxin contamination in durum wheat | Achieved the highest mean recognition percentage of 82.1%. |
| (Ying et al., 2015) | Prediction of early moldy grains (rice, red bean, and oat) | Achieved 93.75 % classification accuracy |
| (Gancarz et al., 2017) | E-nose to predict rapeseed fungal infestation/quality | E-noses were found adequate for assessing fungal infestation. |
| (Xu et al., 2017) | Predicting insect presence in rough rice, infested with different durations | Effectively predicted insect prevalence in stored grain (r- value 0.89). |
| (Mishra et al., 2018a) | Prediction of *Sitophilus granarius* infestation in stored wheat grain using multivariate chemometrics and fuzzy logic | Model efficiently classified the most infested samples and a higher regression co-efficient was found to predict uric acid (0.95) and protein content (0.97) |
| (Mishra et al., 2018b) | Determination of *Rhyzopertha dominica* infestation in wheat using hybrid neuro-fuzzy | Successfully classified various infestation levels. |
| (Nouri et al., 2019) | E-nose was fabricated to detect densities of *Ephestia kuehniella* pest in white flour. | Achieved 90% classification accuracy |
| (Srivastava et al., 2019a) | Fuzzy controller-based E-nose classification of infested rice | Achieved classification accuracy up to 84.75%. |
| (Srivastava et al., 2019b) | Classification of *Rhyzopertha dominica* infestation in stored rice using artificial neural network | The discrimination index was influenced by storage duration and ranged from 88-98%. |
| (Neamsorn et al., 2020) | Designed, fabricated, and evaluated e-nose to classify rancidity and pest damages in brown rice | Satisfactory results for classification of insect damage ($R^2$ =0.98). |
| (Jiarpinijnun et al., 2020) | Early detection of fungal infestation in stored Jasmine brown rice | E-nose coupled with regression model accurately predicted fungal growth ($R^2$=0.969) |
| (Zhou et al., 2021) | Detecting pest-infested rice samples | E-nose detection capabilities were influenced by grain storage duration and insect species. |

| (Leggieri et al., 2022) | Rapid detection of mycotoxin (Deoxynivalenol) in wheat grain samples | Achieved >83% classification accuracy. |
|---|---|---|
| (Zhang et al., 2022) | Identify mold contamination of japonica rice | Excellent performance of the e-nose in identifying japonica rice mildewing |
| (Borowik et al., 2024) | Detecting wheat grain infested by four Fusarium species using e-nose | Achieved classification accuracy of 70%. |
| (Hou et al., 2025) | E-nose was designed to detect and predict densities of *Tribolium castaneum (Herbst)* in stored wheat | Regression models predicted insect density with an R-value of 0.96. |

Apart from pest recognition, the e-nose has demonstrated additional capabilities relevant to grain quality assessment. It can detect a grain sample's age (Zhang & Wang, 2007), monitor changes in grain physical conditions during storage (Deshpandey et al., 2010), and identify signs of grain rancidity (Neamsorn et al., 2020). These additional functions highlight the e-nose's versatility, serving as a multi-purpose tool for post-harvest storage monitoring. Rather than simply serving as a pest detection device, the e-nose can provide comprehensive data on grain quality. This broader capability offers significant advantages to stakeholders by enabling informed decision-making with storage practices, grain handling, and market readiness. Overall, the reviewed literature affirms the effectiveness of e-nose technology in detecting insect infestation, mold growth, and fungal contamination.

The pest species and grain media in which the e-nose was employed are presented in Table 2. The reviewed studies primarily employed e-nose technology in cereal grains media (i.e., wheat and rice), accounting for over 84% of the cases. Undoubtedly, these grains are a staple nutritional source for nearly half of the global population (Divte et al., 2022) and have extended shelf life, thus, stored for extended periods, making them susceptible to biotic pest attacks. In contrast, oilseeds (i.e., rapeseed) and processed grain (i.e., wheat flour) were studied once, indicating a research gap in non-cereal storage systems. Specific insects, commodity, and fungi studied are often dependent on what is common to a particular region, although the highlighted studies cover a good variety of insect species, representing a range of life history characteristics (Table 2).

Infestation by insects has been known to co-occur with fungi or mold, and these contaminants often have a synergistic effect on one another, exacerbating insect or microorganism population growth (Allotey et al., 2001; Mills, 2015). Many insects use pheromones or other chemical communication to communicate with conspecifics, relaying information on food quality, mate location, competition, and oviposition potential (Phillips, 1997). These chemical cues can leave signature VOCs in the environment or directly on the grain or grain-based products, often rendering them unsatisfactory for human use and consumption. External feeders such as the red flour beetle-RFB (*Tribolium castaneum* Herbst), the rusty grain beetle-RGB (*Cryptolestes ferrugineus*), Indian meal moth (*Plodia interpunctella* Hubner), and a variety of mites may be more easily detected as they are often on the surface of grain masses or leave signature chemical cues (USDA, 2016b). However, damage and deterioration of grain can be less conspicuous visually or as odors to humans, especially for insects that live deeper in a grain mass or within a grain kernel. Such internal feeders like rice or maize weevils (*Sitophilus oryzae* or *S. zeamais*), lesser grain borer-LGB (*Rhyzopertha dominica*), or the larger grain borer (*Prostephanus truncatus*) can do a significant amount of damage before they are detected by ordinary measures (Edde et al., 2012). With the aid of an e-nose, early detection of these highly destructive insects could significantly reduce damage and destruction, preserve nutrition of the grains and reduce the impact of subsequent population growth of both insects and microbes. Several studies in the e-nose literature have successfully detected and identified both external feeders and internal feeder insects, further highlighting the superior capability of E-noses over other sensor-based approaches for detecting hidden insect infestations (Table 2).

*Table 2: A summary of e-nose used for various storage pests in different grain medium.*

| Country | Pest | Grain medium | Reference |
| --- | --- | --- | --- |
| Canada | Rusty grain beetle (*Cryptolestes ferrugineus (S.)*), and red flour beetle (*Tribolium castaneum (H.)*) | Wheat | Wu et al (2013) |
| China | Lesser grain borer (*Rhyzopertha dominica (F.)*) | Wheat | Li and Xu (2024) |
| | Lesser grain borer (*Rhyzopertha dominica (F.)*), Grain beetle | Wheat | Zhang & Wang (2007) |
| | Red flour beetle (*Tribolium castaneum (H.)*) | Wheat | Hou et al (2025) |
| | Fungus- mold | Rice, red bean, oat | Ying et al (2015) |
| | Fungus- mold | Japonica rice | Zhang et al (2022) |
| | Red flour beetle (*Tribolium castaneum Herbst*) | Rough rice | Xu et al (2017) |
| India | Rice weevil (*Sitophilus oryzae*) | Rice | Srivastava et al (2019) |
| | Wheat weevil (*Sitophilus granarius*) | Wheat | Mishra et al (2018) |
| | Lesser grain borer (*Rhyzopertha dominica*) | Rice | Srivastava et al (2019) |
| | Lesser grain borer (*Rhyzopertha dominica*) | Wheat | Mishra et al (2018) |
| Iran | Mediterranean flour moth (*Ephestia kuehniella*) | White flour | Nouri et al (2019) |
| Italy | Fungus- mycotoxin (Deoxynivalenol) | Wheat | Leggieri et al (2022) |
| | Fungus- mycotoxin (Deoxynivalenol) | Durum wheat | Lippolis et al (2014) |
| Poland | Fungus- mold | Rapeseed | Gancarz et al (2017) |
| | Fungus (*F. avenaceum, F. langsethiae, F. poae, and F. sporotrichioides*) | Wheat | Borowik et al (2024) |
| Thailand | Red flour beetle (*Tribolium castaneum (H.)*) | Brown rice | Neamsorn et al (2020) |
| | Fungus (*Aspergillus*) | Jasmine brown rice | Jiarpinijnun et al (2020) |
| UK | Flour mites (*Acarus siro L*) | Wheat | Ridgway et al (1999) |
| USA | Rice weevil (*Sitophilus oryzae*), lesser grain borer (*Rhyzopertha dominica*), and red flour beetle (*Tribolium castaneum*) | Rice | Zhou et al (2021) |

The e-nose has been employed across a variety of tasks involving different pest species, grain types, and diverse storage environments and conditions. This collectively demonstrates its adaptability and broader applicability across diverse post-harvest supply chain systems to manage pests and monitor grain parameters. This is particularly advantageous as insects and microbes can be highly influenced by environmental variability and commodity (Gerken & Campbell, 2022; Sinha et al., 1986) and thus, adaptable tools are necessary for implementation across many storage environments.

Table 3 provides an overview of various e-nose systems used in grain storage pest literature, including sensor types, models, and manufacturers. A significantly large number of studies (around 76%) employed commercially available e-nose technology. The well-known and notable manufacturers were Alpha MOS (France) and Airsense Analytics (Germany), which are primarily located in Europe. The commercially available models are patented and developed to analyze generic odors and VOCs, which seem to be working well for storage biotic pest detection tasks (Table 1).

A recent advancement in sensor technology and reduced sensor costs has sparked interest in customized and task-specific e-nose development efforts. Thus, from 2017 to 2025, several researchers designed and fabricated e-noses specific to grain pest detection tasks (Borowik et al., 2024; Gancarz et al., 2017; Hou et al., 2025; Neamsorn et al., 2020; Nouri et al., 2019), achieving performance ranging from acceptable to excellent (Table 1).

Metal Oxide Semiconductors (MOS) are the most widely used sensor type in e-nose. Thus, over 90% of studies (i.e., 19 studies) employed MOS-type e-noses, including both commercial and fabricated e-nose systems (Table 3). The other two studies used conducting polymer-based e-nose systems. Overall, MOS sensors appear to be more effective compared to conducting polymer sensors (Tan & Xu, 2020). The number of MOS sensors varied with the e-nose model type, ranging from 4 to 18. Supplementary Table 4 provides additional details on each MOS sensor type, along with the gases or VOCs they are sensitive to, as used in the respective studies.

*Table 3: Overview of E-nose systems employed in grain storage pest literature.*

| E-nose type | Model | Sensor type | Manufacturer | Reference |
|---|---|---|---|---|
| Commercial | Bloodhound BH114 | 14 Conducting Polymer | Bloodhound Sensors Ltd, Leeds, UK | (Ridgway et al., 1999) |
| Commercial | PEN2 | 10 Metal Oxide Semiconductor (MOS) | Airsense Analytics GmbH, Schwerin, Germany | (Zhang & Wang, 2007) |
| Commercial | Alpha Fox 2000 | 6 MOS | Alpha MOS, Toulouse, France | (Deshpandey et al., 2010) |
| Commercial | AOS ISE Nose 2000 | 12 MOS | SoaTec S.r.l., Parma, Italy | (Lippolis et al., 2014) |
| Commercial | Not specified | 12 MOS | - | (Li & Xu, 2014) |
| Commercial | Not specified | 8 MOS | - | (Ying et al., 2015) |
| Fabricated | - | 8 MOS | - | (Gancarz et al., 2017) |
| Fabricated | - | 6 MOS | - | (Nouri et al., 2019) |
| Fabricated | - | 8 MOS. | - | (Neamsorn et al., 2020) |
| Commercial | Cyranose-320 | 32 Conducting Polymer | Sensigent, Inc, California, U.S. | (Zhou et al., 2021) |
| Fabricated | - | 4 MOS | - | (Borowik et al., 2024) |
| Fabricated | - | 10 MOS | - | (Hou et al., 2025) |
| Commercial | Alpha MOS FOX-3000 | 12 MOS | Alpha MOS, Toulouse, France | (Jiarpinijnun et al., 2020; Wu et al., 2013) |
| Commercial | Fox 4000 | 18 MOS | Alpha MOS, Toulouse, France | (Mishra et al., 2018a; Mishra et al., 2018b) |
| Commercial | Alpha Soft Fox 2.0 | 18 MOS | Alpha MOS, Toulouse, France | (Srivastava et al., 2019a, 2019b) |
| Commercial | PEN3 | 10 MOS | Airsense Analytics Inc., Germany | (Leggieri et al., 2022; Xu et al., 2017; Zhang et al., 2022) |

The e-nose produces digital signals that require analysis and interpretation to obtain meaningful information. These digital signals are proportional to the number of sensors (i.e., MOS or CP) used in the e-nose, resulting in a multi-variate and large dataset. Thus, data analysis is a critical part of e-nose operation, which is mainly used for qualitative classification and quantitative prediction. A brief overview of the data analysis approach used in e-nose literature is presented in Table 4. Two learning approaches that are commonly used include supervised learning, which uses human-labeled data to train prediction models, and unsupervised learning, which discovers patterns in unlabeled data. Principal component analysis (PCA) and linear discriminant analysis (LDA) are the most used methods for e-nose data processing (Table 4). Both PCA and LDA are dimensionality reduction techniques that simplify large datasets into smaller sets while preserving significant patterns or information commonly used for classification and feature selection tasks. Other multivariate data analysis techniques employed in e-nose data analysis include support vector machines, cluster analysis (i.e., Fuzzy C-Mean, hierarchical, K-means), and neural networks (i.e., olfactory, probabilistic, back-propagation). Several studies reported that the machine learning-based approach seems to outperform the conventional statistical approach. For example, SVM outperformed LDA in the classification task of fungal infestation in brown rice (Jiarpinijnun et al., 2020) and wheat samples (Borowik et al., 2024), while the probabilistic neural network outperformed the statistical approach in classifying early moldy grain (Ying et al., 2015) and pest infestation in stored rice (Srivastava et al., 2019b). Moreover, the combined olfactory neural network and SVM approach delivered 100% classification accuracy in insect infestation in wheat (Li & Xu, 2014).

Apart from qualitative analysis, several researchers used a regression approach for quantitative analysis or prediction. For example, partial least squares (PLS) regression was developed to predict fungal growth in brown rice with an $R^2$ of 0.97 (Jiarpinijnun et al., 2020), while the Gaussian process regression predicted red flour beetle densities in wheat with a fitting coefficient of 0.96 (Hou et al., 2025). These results indicate that e-nose can be used for quantitative analysis to estimate pest densities or threshold estimation often required to establish pest treatment decisions.

E-nose can contain up to 18 MOS sensors, resulting in 18 individual VOC response signals. However, not all sensor responses will be equally important, and response strength will vary based on VOC/gas concentration in sample space. For example, Hou et al., (2025) reported that all 10 MOS sensors responded differently to volatile odors produced by red flour beetle infestation in wheat, and weaker sensors did not contribute significantly to the overall signal response. This is primarily because the broad-spectrum response and cross-sensitivity of the sensor array result in the generation of redundant information (Hou et al., 2025). This less informative or redundant sensor response will influence the model prediction ability. Hence, several researchers focused on optimizing sensor arrays to improve model accuracy. A hybrid modeling technique, called an adaptive neuro-fuzzy system, was developed to optimize the 18-MOS sensor array, where only two sensor responses were found to be sensitive to detect pest infection in wheat (Mishra et al., 2018b). Fuzzy logic was used to screen relatively sensitive sensors toward pest infestation in wheat (Mishra et al., 2018a) and rice (Srivastava et al., 2019a), considering the response of four and six sensors, respectively, out of 18 MOS sensors. Additionally, multivariate statistical methods were used to optimize sensor arrays by reducing the number of MOS sensors from 10 to 6, resulting in a 60% reduction in features (Hou et al., 2025). Similarly, an e-nose equipped with a conducting polymer sensor type, sensor selection were recommended during the model training process to improve identification accuracy (Zhou et al., 2021). In a nutshell, reducing less relevant feature information has shown significant improvement in model prediction ability (Hou et al., 2025; Mishra et al., 2018a).

*Table 4: Overview of the learning approaches used in data analysis in e-nose.*

| Learning approach | Approach types | Reference |
|---|---|---|
| Unsupervised learning methods | Principle component analysis (PCA) | (Deshpandey et al., 2010; Gancarz et al., 2017; Mishra et al., 2018a; Mishra et al., 2018b; Neamsorn et al., 2020; Nouri et al., 2019; Srivastava et al., 2019b; Wu et al., 2013; Ying et al., 2015; Zhang et al., 2022; Zhou et al., 2021) |
| | Fuzzy C-Means clustering | (Xu et al., 2017) |
| Supervised learning methods | Combined olfactory neural network | (Li & Xu, 2014) |
| | Support vector machines (SVM) | (Borowik et al., 2024; Jiarpinijnun et al., 2020; Li & Xu, 2014) |
| | Probabilistic neural network | (Srivastava et al., 2019b; Ying et al., 2015) |
| | Back propagation neural network | (Srivastava et al., 2019b; Xu et al., 2017; Ying et al., 2015) |
| | Linear discriminant analysis (LDA) | (Borowik et al., 2024; Gancarz et al., 2017; Jiarpinijnun et al., 2020; Neamsorn et al., 2020; Nouri et al., 2019; Ridgway et al., 1999; Xu et al., 2017). |
| | Regression analysis | (Hou et al., 2025) |
| | Classification and regression trees | (Leggieri et al., 2022) |
| | Discriminate function analysis | (Deshpandey et al., 2010; Lippolis et al., 2014; Srivastava et al., 2019b). |
| | Canonical discriminant analysis | (Zhou et al., 2021) |
| | Discriminant factorial analysis | (Wu et al., 2013) |
| | Partial least square analysis | (Jiarpinijnun et al., 2020; Wu et al., 2013) |
| | Hierarchical cluster analysis | (Mishra et al., 2018a; Zhang et al., 2022). |
| | Soft Independent Modeling by Class Analogy | (Srivastava et al., 2019b) |
| | K-means clustering | (Xu et al., 2017) |
| Hybrid and advanced techniques | Adaptive neuro-fuzzy system | (Mishra et al., 2018a) |
| | Fuzzy logic | (Mishra et al., 2018a; Srivastava et al., 2019b) |

Cereal grains have extended shelf life and can be stored for several months to a year. However, these grains produce different volatiles with changing storage times. Insect presence, their metabolic activities and excreta produce an off-odor with distinct VOC, further affecting grain quality (zhang & wang 2007). Therefore, e-nose performance was influenced by storage duration as these VOCs undergo biochemical changes over time (Mishra et al., 2018b). For example, Zhang et al., (2022) reported that e-nose response was significantly less for rice storage duration above 10 days for identifying rice mildewing. Likewise,

Zhou et al., (2021) reported detection of infestation in rice using an e-nose only after four weeks of grain storage.

Moreover, as the pest density increases in grain samples, the VOC production is expected to be higher since both pest by-products and grain structural degradation will increase over time (Mishra et al., 2018b). Several studies reported that e-nose is capable of differentiating grain samples with varying pest densities (Lippolis et al., 2014; Nouri et al., 2019; Wu et al., 2013; Xu et al., 2017). However, the sensor response is significantly reduced at lower pest densities. For example, Wu et al., (2013) reported that e-nose was able to differentiate 1 Red Flour Beetle (RFB)/kg infestation level from 20 RFBs/kg infestation level in wheat, but sensor responses were 75% at 1 insect/kg and 338% at 20 insects/kg. Likewise, Xu et al. (2017) suggested that the heavier the pest infestation, the more unique VOCs which leads to better e-nose response for heavy-degree and middle-degree than light-degree infestation of RFB in rough rice.

The e-nose performance appears to be influenced by pest species types. For example, Zhou et al., (2021) reported that the employed e-nose was unable to differentiate between rice samples infested with lesser grain borer, but the same e-nose showed a high sensor response to the VOC produced by rice weevil and red flour beetle. Likewise, Wu et al., (2013) reported that employed e-nose failed to detect the presence of rusty grain beetle in wheat but could detect the presence of red flour beetle. This might be due to the common fact that different pests may exhibit varying levels of odor depending on life stage, genetic makeup, or metabolism. This statement is supported by (Seitz & Sauer, 1996) research which revealed that red flour beetles produce some off-odor while rusty grain beetles produce very little objectionable odor, even when present in large numbers.

Moreover, the e-nose performance was influenced by storage conditions such as temperature and grain storage moisture. Zhou et al. (2021) reported that the employed e-nose had good classification accuracy for clean and infested rice samples stored above 30°C but delivered low accuracy for sample temperature of 15°C for the tested insects. Similarly, Wu et al., (2013) showed that e-nose failed to detect RFB presence in wheat at 18% moisture but was only able to detect RFB presence at high infestation levels at 14% and 16% moisture content. This might be because insect activity and growth are influenced by storage temperature and humidity, impacting their development, survival, and reproduction. Thus, a lower temperature would lead to minimal activity to produce VOC compounds that will get detected by e-nose. Additionally, insect activity, survival, and reproduction do vary with species type. Thus, pest-specific temperature and moisture profiles need to be considered before putting in an e-nose for the pest detection tasks. In summary, e-nose sensors are sensitive to pest species type, pest density, storage temperature, and moisture content. Moreover, the e-nose sensor response will vary based on storage duration, grain type, and storage environment. Therefore, it is possible that the same e-nose configuration used at different locations, storage types, or structures may produce different outputs since the above-mentioned factors vary from place to place. Thus, the e-nose will need calibration to adjust to local crops, pests, and climate conditions.

Most of the e-nose literature included in this study either employed or experimentally validated its performance on a small scale (i.e., sample jar) in a controlled laboratory setup with a few selected variables. While controlled conditions allow precise calibration and repeatability, they do not fully represent the complexities of real-world storage scenarios. The literature on field scale-level (e.g., grain bin, storage bag) validation is scarce and expanding such literature can be critical to enhance the feasibility, reliability, and eventual adoption of e-nose technology by stakeholders. To deploy the e-nose in an actual grain bin, Wu et al. (2013) highlighted the need for a specialized gas sampling system capable of extracting VOC samples from multiple bin locations to the e-nose for comprehensive monitoring. However, this may not be economical for small-scale grain storage since the cost of data analysis would be significant (Wu et al., 2013).

Moreover, most studies employed generic or commercially available e-nose systems, which perform well for pest detection tasks. However, these e-noses equipped with multiple MOS or CP sensors produce large multivariate response data, which results in poor model analysis and classification performance (Hou et al., 2025; Mishra et al., 2018b). Thus, sensor optimization, using expertise in data processing, needs to be performed with an advanced modeling approach to obtain good detection performance (Hou et al., 2025; Mishra et al., 2018b). Presently, e-nose operation is not fully automated but requires familiarity with e-nose operation, experience, and training in sample preparation, gas sensing, sensor selection, data processing, and analysis to obtain useable results (Wilson & Baietto, 2009). E-nose sensors are sensitive to grain moisture and temperature, which can result in inaccurate results (Wilson & Baietto, 2009; Wu et al., 2013; Zhou et al., 2021). Additionally, oxidation can damage sensor material coating which needs to be replaced on a yearly basis, requiring periodic maintenance and cost (Wu et al., 2013). Some of the other disadvantages of e-nose operation include sensor drift, poor repeatability, and comparability (Tan & Xu, 2020; Wilson & Baietto, 2009). However, ongoing advancements in new sensor technology and data processing techniques will provide continuous improvements in e-nose technology.

Despite the challenges, e-nose technology has emerged as a safer, cost-effective, rapid, non-invasive, and accurate solution for pest detection in stored grain environments. The major significant advantage of e-nose is to detect and identify both microscopic organisms (e,g., fungi), and hidden insects (i.e., presence and infestations), which are often missed by other sensor-based methods (e.g., imaging, acoustics, or NIR spectroscopy). The e-nose can provide both qualitative (i.e., classification) and quantitative (i.e., pest infestation level and pest density) information, which are quite challenging to obtain with other sensor-based methods. This capability enables early warning and supports timely and data-driven decision-making for IPM strategies (Hagstrum & Athanassiou, 2019). Remarkably, e-nose can detect pest presence at very low infestation levels (e.g., 1 insect/kg grain), which stakeholders can use to avoid significant economic losses due to commodity devaluation, as just two live insects can lead to price discounts for grain sellers (Hagstrum, 2012). Furthermore, e-nose applications have expanded beyond pest detection to include grain quality assessments, such as age determination, grain physical properties, and rancidity detection, emphasizing its multifunctional utility. Moreover, integrating e-nose technology with other sensor-based pest detection methods (i.e., acoustic and camera-based systems) can improve the overall accuracy and reliability of pest detection. The complementary strengths of e-noses can help overcome the limitations of individual sensor-based approaches. Due to its rapid and accurate performance, e-noses also show promise as a diagnostic tool at point-of-entry locations for cross-border inspection, helping prevent the spread of invasive pest species to new geographical regions.

**Conclusions:**

In this study, we adopted a systematic literature review methodology to critically evaluate all relevant research on e-nose application stored grain pest detection, identification, and monitoring. This review systematically documents and presents a well-curated collection of reference literature on E-noses, which will be particularly valuable to stored product entomologists, engineers, and researchers working in this domain. The following conclusion can be drawn from this review:

- E-nose technology is low-cost, rapid, non-invasive, and accurate for pest detection.
- E-nose can detect and identify microscopic (Fungi) and hidden insects with good accuracy.
- E-nose can also be used for grain quality assessments (e.g., age determination and rancidity), emphasizing its multifunctional utility.
- E-nose is an inexpensive and portable tool that would allow off-site grain inspection.
- Metal Oxide Semiconductor (MOS) sensors are the most widely used sensor type in both commercial and fabricated E-nose systems, primarily due to their sensitivity, affordability, and robustness.
- Cereal grains such as wheat and rice were most studied, while the literature on oilseeds and finished products is very limited.
- A noticeable trend of custom e-nose development has been observed, which is focusing more on task-specific to stored grain pest detection.
- A wide range of learning methods were employed for data processing, pattern recognition, and classification. However, PCA, LDA, SVM, and neural network-based models were among the most frequently applied techniques.
- E-nose performance is influenced by storage duration, storage parameters (temperature, moisture), pest species type, and pest density.
- The major challenges include sensor array optimization or selection, large data processing, poor repeatability, and comparability among measurements.
- The collective advantage of the e-nose system can overcome the limitations of existing sensor-based detection systems and can serve as an alternative tool for pest detection, which has the potential to reduce overall food-grain losses and economic costs via timely actions.

*Table 5: Overview of E-nose systems with detailed on MOS sensor type employed in grain storage pest literature.*

| Employed E-nose sensor type | Details MOS sensor types and sensitive/reactive gas | Reference |
|---|---|---|
| Commercial E-nose with 12 MOS. | Not mentioned | Li and Xu (2024) |
| Commercial e-nose with 8 MOS. | TGS-825: Sulfide; TGS-821: Flammable gases; TGS-826: Ammonia gas; TGS-822: Ethanol, aromatic hydrocarbons; TGS-842: Hydrocarbon component gas; TGS-813: Methane, propane, butane; TGS-2610: Propane, butane; TGS-2201: Nitrogen oxides. | Ying et al., (2015) |
| Commercial e-nose, model- Bloodhound BH114 (Bloodhound Sensors Ltd, Leeds, UK), with 14 conducting polymer sensors. | Not mentioned | Ridgway et al., (1999) |
| Commercial e-nose, model- Alpha MOS FOX-3000 (Alpha MOS, Toulouse, France), equipped with 12 MOS. | LY2/LG: Oxidizing gas; LY2/G: Ammonia/carbon monoxide; LY2/AA: ethanol; LY2/GH: Ammonia/organic amine; LY2/gCTL: Hydrogen sulfide; LY2/gCT: Propane/butane; T30/1: Organic solvents; P10/1: Hydrocarbons; P10/2: Methane; P40/1: Fluorine, T70/2: Aromatic compounds; PA/2: ethanol/ammonia/organic ammine | Wu et al (2013), Jiarpinijnun et al., (2020) |
| Commercial e-nose, model- Alpha Soft Fox 2.0 equipped with 18 MOS. | LY2/LG: Fluorine, chloride, oxynitride, sulfide; LY2/G: Carbon, oxygen; LY2/AA: Alcohol, acetone, ammonia, aldehydes; LY2/GH Ammonia, amines; LY2/gCT1: Hydrogen sulfide; LY2/gCT: Propane, butane; T30/1: Polar compounds, hydrogen chloride, aldehydes, ketones, hexanoic acid, uric acid; P10/1: Non-polar compounds, methane, ethane; P10/2: Aldehydes, ketones, hexanoic acid, pentanoic acid; P40/1: Fluorine, chlorine, acids; T70/2: Toluene, xylene, carbon monoxide, carbon compounds; PA/2: Ethanol, ammonia, amine compounds; P30/1: Hydrocarbons, ammonia, ethanol, amines; P40/2: Chlorine, hydrogen sulfides, fluorine; P30/2: Hydrogen sulfides, ketones; T40/2 Chlorine, carbon compounds, uric acid; T40/1: Fluorine, uric acid; TA/2: Ethanol | Srivastava et al., (2019a), Srivastava et al., (2019b), |
| Commercial e-nose, model- Artificial Olfactory System ISE Nose 2000 (SoaTec S.r.l., Parma, Italy) equipped with 12 MOS. | Not mentioned | Lippolis et al., (2014) |
| Commercial e-nose, model: Alpha Fox 2000 (Alpha MOS, Toulouse, France) equipped with 6 MOS. | P30/1: Organic solvents and polar molecules; P10/1: hydrocarbons; P10/2: methane, propane and other non-polar molecules; P40/2 aldehydes; SX21: aldehydes, rancid smells, long-chain fatty acids; T70/2: alcoholic vapors | Deshpande et al., (2010) |
| Commercial e-nose, model: PEN3 E-nose (Airsense Analytics Inc., Germany) with 10 MOS. | W1C: aromatic; W5S: broad range; W3C: aromatic; W6S: hydrogen; W5C: aromatic-aliphatic; W1S: broad-methane; W1W: sulfur-organic; W2S: broad-alcohol; W2W: sulfur-chlorinate and W3S: methane-aliphatic. | Xu et al (2017), Leggieri et al., (2022), Zhang et al., (2022) |

Supplementary material

| Device | Sensors | Reference |
|---|---|---|
| Commercial e-nose, model-Fox 4000 (Alpha MOS, Toulouse, France) equipped with 18 MOS. | LY2/LG: Fluorine, chloride, oxynitride, sulfide; LY2/G: Carbon, oxygen; LY2/AA: Alcohol, acetone, ammonia, aldehydes; LY2/GH Ammonia, amines; LY2/gCT1: Hydrogen sulfide; LY2/gCT: Propane, butane; T30/1: Polar compounds, hydrogen chloride, aldehydes, ketones, hexanoic acid, uric acid; P10/1: Non-polar compounds, methane, ethane; P10/2: Aldehydes, ketones, hexanoic acid, pentanoic acid; P40/1: Fluorine, chlorine, acids; T70/2: Toluene, xylene, carbon monoxide, carbon compounds; PA/2: Ethanol, ammonia, amine compounds; P30/1: Hydrocarbons, ammonia, ethanol, amines; P40/2: Chlorine, hydrogen sulfides, fluorine; P30/2: Hydrogen sulfides, ketones; T40/2 Chlorine, carbon compounds, uric acid; T40/1: Fluorine, uric acid; TA/2: Ethanol | Mishra et al., (2018a), Mishra et al., (2018a) |
| Commercial e-nose, model-Cyranose-320 (Sensigent, Inc, California) consisting of 32 carbon polymer sensors | | Zhou et al., (2021) |
| Commercial E-nose, model-PEN2 (Airsense Analytics GmbH, Schwerin, Germany) with 10 MOS. | MOS-1: Aromatic; MOS-2: broad-range; MOS-3: aromatic; MOS-4: hydrogen; MOS-5: arom-aliph; MOS-6: broad-methane; MOS-7: sulfur-organic; MOS-8: broad alcohol; MOS-9: sulph-chlor; MOS-10: methane-aliph. | Zhang & Wang (2007) |
| Fabricated e-nose with 10 MOS. | TGS2620: Ethanol, Hydrogen; TGS813: Hydrogen, Isobutane, Propane; TGS2600: Ammonia, Ethanol, Isobutane; MQ135: Ammonia, Hydrogen sulfide, Benzene; TGS2603: Trimethylamine, Hydrogen sulfide; MQ137: Ammonia, Ethanol; WSP2110: Benzene, Formaldehyde; MQ138: Benzene, Acetone; TGS2610: Propane, Butane; TGS822: Acetone, Ethanol, Benzene. | Hou et al., (2025) |
| Fabricated e-nose with 4 MOS. | TGS 2603: sulfurous odors from waste materials or spoiled food; TGS 2610: alcohol and LP gas; TGS 2611: alcohol and methane gas; TGS 2620: organic solvents and other volatile vapors. | Borowik et al., (2024) |
| Fabricated e-nose with 6 MOS. | MQ-2: LPG, Propane and Hydrogen; MQ-3: Alcohol vapor; MQ-4: CH4, Natural gas; MQ-5: LPG, methane, coal gas; MQ-6: LPG, ISO-butane, propane; MQ-8: Hydrogen. | Nouri et al., (2019) |
| Fabricated e-nose with 8 MOS. | Organic solvent vapor, ammonia, air contaminants, butane, propane, alcohol, solvent vapor, alcohol, carbon monoxide, hydrogen | Neamsorn et al., (2020) |
| Fabricated with 8 MOS. | TGS2600-B00: General air contaminants, hydrogen, and carbon monoxide; TGS2610-C00: LP gas, butane; TGS2602-B00: Ammonia, Hydrogen sulfide; TGS2611-C00: Natural gas, methane; TGS2611-E00: Natural gas, methane (carbon filter); TGS2612-D00: Propane, methane, solvent vapors, iso-butane (carbon filter); TGS2620-C00: Solvent vapors, volatile vapors, alcohol; AS-MLV-P2: CO, butane, methane, ethanol, hydrogen. | Gancarz et al., (2017) |